\documentclass[11pt,a4paper]{article}
\usepackage{graphicx}

\title{\flushleft \textbf{Monte Carlo simulation of melting transition on DNA nanocompartment}}
\date{}

\begin{document}
\maketitle{}
\begin{quotation}
\noindent $\textrm{Song Chang}^{1,\,2}$, $\textrm{Youdong Mao}^{1}$,
$\textrm{Zhengwei Xie}^{2}$
$\textrm{Chunxiong Luo}^{1}$, $\textrm{Qi Ouyang}^{1,\,2,\,3}$\\
$\textrm{}^{1}$Laboratory of Biotechnology, School of Physics,
Peking University, Beijing 100871, China\\
$\textrm{}^{2}$Center for Theoretical Biology, Peking University,
Beijing 100871, China\\
$\textrm{}^{3}$To whom correspondence should be addressed.\\
E-mail: qi@pku.edu.cn\\
\end{quotation}

\begin{abstract}
\noindent DNA nanocompartment is a typical DNA-based machine whose
function is dependent of molecular collective effect. Fundamental
properties of the device have been addressed via electrochemical
analysis, fluorescent microscopy, and atomic force microscopy.
Interesting and novel phenomena emerged during the switching of
the device. We have found that DNAs in this system exhibit a much
steep melting transition compared to ones in bulk solution or
conventional DNA array. To achieve an understanding to this
discrepancy, we introduced DNA-DNA interaction potential to the
conventional Ising-like Zimm-Bragg theory and Peyrard-Bishop model
of DNA melting. To avoid unrealistic numerical calculation caused
by modification of the Peyrard-Bishop nonlinear Hamiltonian with
the DNA-DNA interaction, we established coarse-gained Monte Carlo
recursion relations by elucidation of five components of energy
change during melting transition. The result suggests that DNA-DNA
interaction potential accounts for the observed steep transition.\\
\end{abstract}

\noindent\textbf{1.Introduction}\\
\noindent Studies on the physical chemistry of DNA denaturation
have been lasted for almost forty years [1-3]. In 1964, Lifson
proposed that a phase transition exists in one-dimensional polymer
structure. He introduced several pivotal concepts, like sequence
partition function, sequence generating function, etc., and
established a systematic method to calculate the partition
function [1]. These allow us to derive important thermodynamic
quantities of the system. In 1966, Poland and Scherage applied
Lifson's method to conduct research on amino acid and nucleic acid
chains. They built Poland-Scherage (PS) model for calculating the
sequence partition function and discussing the behavior of
polymers in melting transitions.\\
\indent Another excellent progress would be the building of
Peyrard-Bishop (PB) model [4,5] for DNA chains. In PB model, the
Hamiltonian of a single DNA chain, which is constructed by phonon
calculations, is given so that we can obtain the system properties
through statistical physics method. The PB model has introduced
mathematical formula of stacking energy, as well as the kinetic
energy and potential energy of each base pair. By theoretical
calculation, one can show the entropy-driven transition that leads
DNA to shift from ordered state to disorder one [6,7].\\
\indent However, all these works have not involved the DNA-DNA
interactions because the subject investigated is DNAs in bulk
solution, and the interaction between them has ever been
neglected. The main idea of this paper is to inspect the influence
of collective effect on the DNA melting process, primarily
motivated by the experiment results of DNA nanocompartment [8,9].
Under the enlightenment of Poland-Scherage model and Zimm-Bragg
model [10], we simplify Peyrard-Bishop model to meet a reasonable
Monte Carlo simulation by the elucidation of five components of
energy changes during melting transition. The result shows that
the melting temperature and transition duration depend on whether
we take into account the DNA-DNA interactions among columnar
assemblies of DNA.\\

\noindent\textbf{2.Experiment}\\
\noindent Recently, we found that specially designed DNA array can
form a molecular cage on surfaces [8,9]. This molecular cage is
switchable due to allosteric transformation driven by the collective
hybridization of DNA. We named it "active DNA nanocompartment
(ADNC)". Typical DNA motif designed to fabricate ADNC comprises two
contiguous elements (inset to figure 1a): a double-stranded DNA
(dsDNA) whose array is responsible for a compact membrane (figure
1a, right), and a single-stranded DNA (ssDNA) serving as skeleton
supporting the dsDNA membrane, which is terminated on its 5¡ä end by
a surface linker such as an alkanethiol group that can be tethered
to gold surface with a sulphur-gold bond [9] or an amino group that
can be tethered to $\textrm{SiO}_{2}$ substrate with specific
surface attachment chemistry [11]. Because the diameter of ssDNA is
much smaller than that of dsDNA, a compartment with designable
effective height (heff, $5\sim50 nm$, commensurate with the length
of ssDNA skeleton) can form between the dsDNA membrane and substrate
surface.\\
\indent Since ADNC is reversibly switchable, it is able to encage
molecules with suitable size. We name this phenomenon molecular
encaging effect. Both electrochemical methods [12] and fluorescent
microscopy are used to substantiate the molecular encaging effect
and the reversibility of switching. Once the closed ADNC entraps
some chemical reporters, the surface concentration ($\Gamma_{nc}$)
of the encaged reporters can be determined by cyclic voltammetry
or fluorescent microscopy. Figure 1b shows the isotherms of the
molecular encaging effect for fluorescein
($\textrm{C}_{20}\textrm{H}_{10}\textrm{Na}_{2}\textrm{O}_{5}$).
Figure 1c presents the melting curves of ADNC. Using the encaged
molecules as indicator greatly sharpens the melting profiles for
the perfectly complementary targets, and flattens denaturation
profiles for the strands with a wobble mismatch. The observation
shows that single-base mismatched strands are incapable of closing
ADNC on surfaces. The result is highly consistent to our
observation by electrochemical analysis [12]. These observations
bring up an intriguing question: why the melting curves exhibit so
steep transition compared to the case of DNA in bulk solutions or
on a loosely packed microarray? We try to address this question in
this paper.\\
\indent Worthy of mention is that the steepness of melting
transition is useful when the ADNC is applied to DNA detection
[8,9]. First, it greatly enhances the discrepancy of perfect
targets and single mismatches. This provides much enhanced
specificity in DNA recognition, $100:1\sim105:1$ of our system
versus $2.7:1$ of conventional system. Second, more sensitivity is
obtained with optimally decreased ambiguity. Therefore, the
clarification of the origin of the steep shape should help us to
further extend the experience to related fields or generate new
techniques.\\
\begin{figure}[!hp]
\begin{center}
\includegraphics[width=\textwidth]{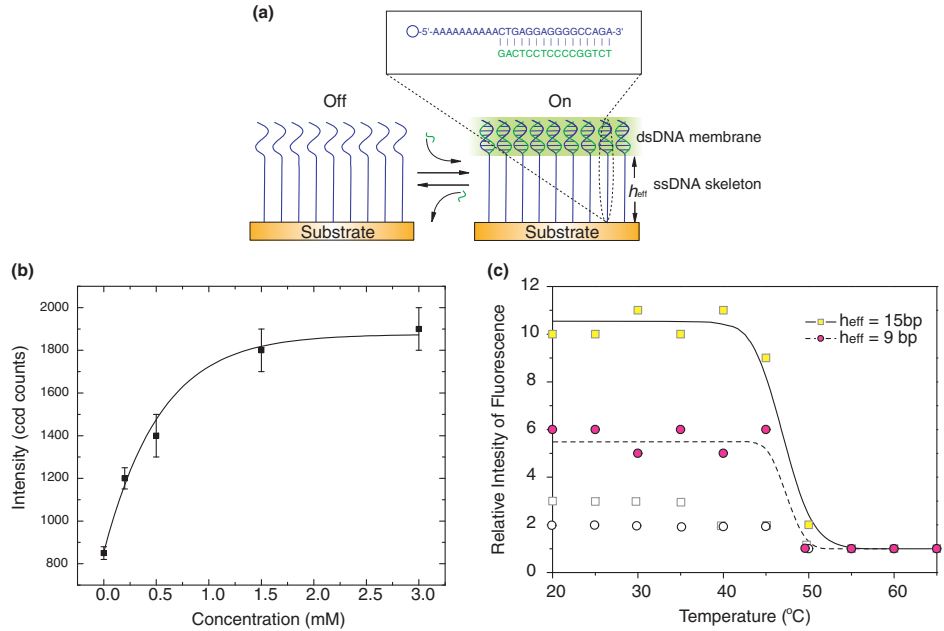}
\caption{ (a) Schematic drawing of a dual-state ADNC. By adding or
removing 'fuel' strands (shorter segment), the ADNC can be switched
between on (right) and off (left) state. Inset, a typical sequence
used to fabricate ADNC. The 'fuel' strands is a segment of human p53
gene containing one site of most frequent mutation. (b) Isotherms of
$\Gamma$ for fluorescein encaged in a closed ADNC ($h_{eff} =
15bp$). The isotherm fits well to the Langmuir model: $x/\Gamma_{nc}
= (1/\Gamma_{nc,max})x + (1/K\Gamma_{nc,max})$, where $x$ is the
concentration of the reporter and $K$ is the association constant
per site, $\Gamma_{nc}$ the surface concentration of encaged
molecules. (c) Melting curves using the encaged fluorescein
molecules as indicators. Filled circles or squares are corresponding
to perfect complementary strands, and hollow circles or squares to
single-base mismatched strands. The unit of relative intensity of
fluorescein is defined as the light intensity of $5 \mu l\,10nM$
fluorescein on a spot size with $5 mm$ diameter. Inset shows the
length of the ssDNA skeleton of nanocompartment and the
complementary type. The background noise is within $1$ unit.}
\end{center}
\end{figure}

\noindent\textbf{3.Modeling}\\
\noindent Taking into account the directional specificity of the
hydrogen bonds, the Hamiltonian of a single DNA chain is obtained
as following form according to PB model [4-6],
\begin{equation}
H_y = \sum_{n} \bigg[ \frac{1}{2}m\dot{y_n}^2 + w(y_n, y_{n-1}) +
V(y_n)\bigg]
\end{equation}
\noindent where the $y_n$ is the component of the relative
displacement of bases along the direction of hydrogen bond. The
stacking energy $w(y_n, y_{n-1})$ corresponds to the interaction
between neighboring base pair in one DNA chain
\begin{equation}
w(y_n, y_{n-1}) = \frac{k}{2} \bigg[1+ \rho
e^{-\alpha(y_n+y_{n-1})}\bigg](y_n-y_{n-1})^2
\end{equation}
\indent The Morse potential  describes the potential for the
hydrogen bonds
\begin{equation}
V=D(e^{-\alpha y}-1)^2
\end{equation}
\indent However, in this study, the Hamiltonian in equation (1) is
not sufficient; it neglects the structure of close-packing of DNA
in ADNC. In our system, one should take into account the
interactions between the nearest neighboring molecules [13,14]. To
model the interaction, one envisions the molecules as rigid
cylinders, carrying helical and continuous line charges on their
surfaces. Each DNA duplex carries the negative charge of
phosphates plus a compensating positive charge from the adsorbed
counterions. Let $0 < \theta < 1$ be the degree of charge
compensation, $f_1$, $f_2$ and $f_3$ the fractions of condensed
counterions in the minor and major grooves ($f_1 + f_2 + f_3 =
1$). The mobile counterions in solution screen the Coulomb
interactions between the two molecules, causing at large
separations an exponential decay of the latter with the Debye
screening length $\kappa^{-1}$. The solvent is accounted for by
its dielectric constant $\varepsilon$. The structural parameters
of B-DNA are half azimuthal width of the minor groove
$\tilde{\phi_s}\approx0.4\pi$ , pitch $H\approx 34\textrm{\AA}$($g
= 2\pi/H$), and hard-core radius $a = 9\textrm{\AA}$. We take the
following form for the pair interaction potential [15-18]:
\begin{eqnarray}
u(R,\phi) &=& u_{0} \sum_{n=-\infty}^{\infty} \bigg[ f_1 \theta +
(-1)^n f_2 \theta -(1-f_3 \theta)
\cos(n\tilde{\phi_s})\bigg]^2\nonumber\\
&&\times \frac{(-1)^n\cos(ng\Delta z)K_0 (\kappa_n
R)-\Omega_{n,n}(\kappa_nR,\kappa_na)}{(\kappa_n/\kappa)^2[K'_n(\kappa_na)]^2}
\end{eqnarray}
\noindent where $R(>2a)$ is the distance between the two parallel
DNA molecules, $\Delta z$ a vertical displacement, equivalent to a
''spin angle'' $\phi= gz$. Here, $u_0 = 8\pi \sigma^2/\varepsilon
\kappa^2$ (about $2.9k_BT/\textrm{\AA}$ at physiological ionic
strength), and $\kappa_n = \sqrt{\kappa^2+n^2g^2}$.
$\Omega_{n,m}(x,y)$ is given by
\begin{equation}
\Omega_{n,m}(x,y) =
\sum_{j=-\infty}^{\infty}\bigg[K_{n-j}(x)K_{j-m}(y)\frac{I'_j(y)}{K'_j(y)}
\bigg]
\end{equation}
\noindent with the modified Bessel functions $K_n(x)$ and
$I_j(y)$. The primes denote derivatives. The sum rapidly
converges, and it can be truncated after $|n| = 2$. Since
$\kappa_nR > 3$ and $g\sim\kappa$, each of the terms in the sum
decreases exponentially at increasing $R$ with the
decay length $\kappa_n^{-1}\propto1/n$.\\
\indent Figure 2 present a scheme of interaction between two
neighboring columnar DNA molecules charged with counterions on its
surface. The distance between two DNA columns in our simulation is
about $30\textrm{\AA}$ and the helical pitch of DNA molecule is
about $36\textrm{\AA}$. For brevity, we take the mean-field
approximation that the pair
interactions mainly exist between charges in the same height.\\
\begin{figure}[!hp]
\begin{center}
\includegraphics[width=\textwidth]{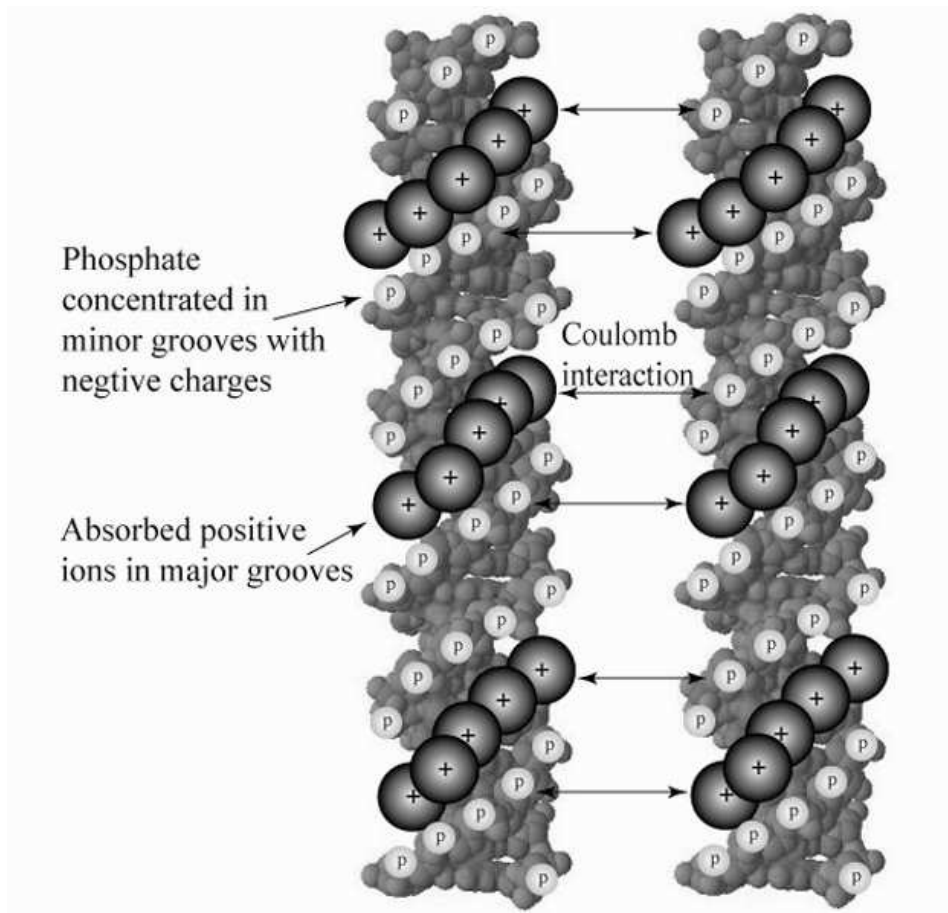}
\caption{The pair interaction between two parallel B-DNA double
helixes. The black balls with positive signs in the center represent
the absorbed positive counterions, while the little grey balls
represent the phosphate carrying negative charges. Each DNA duplex
carries the negative charge of phosphates with area density of $16.8
\mu C/cm^2$ plus a compensating positive charge coming from the
adsorbed counterions. We take the assumption that, and the distance
between them are approximately $30\textrm{\AA}$. }
\end{center}
\end{figure}

\noindent\textbf{4.Monte Carlo Simulation}\\
\noindent Let $t$ be the dimensionless variable to mark the time
series of simulation $(t = 0, 1, 2\ldots)$ and $T$ the
environmental temperature. Assuming that $M\times N$ DNAs are on
the ADNC, the position of each DNA can be represented by its
coordinates $(x, y)$, where $x,y\in N$, and $0<x<M, 0<y<N$. All
DNA molecules in ADNC have identical sequence with $P$ base pairs.
Therefore there is the collection of $M\times N\times P$ base
pairs. The degree of freedom of the system is also $M\times
N\times P$. The position of each base pair is thus represented by
coordinates $(x, y, i)$, where $i\in N, 0<i<P$. We take that the
indices of base pairs is assigned from the bottom to the top
of the DNA.\\
\indent At the time $t_0$, the state for an arbitrary base pair at
$(x_0, y_0, i)$ with well-formed hydrogen bonds is represented as
$(x_0, y_0, i, t_0) =1$. Contrarily, the state of a base pair with
decoupled hydrogen bonds is denoted as  $(x_0, y_0, i, t_0) =0$
[10, 19]. $\psi(x, y, i, t)$ is a function of the time and the
position of the base pair. Therefore, the state of each DNA
molecules in ADNC can be represented by a sequence of digits. The
number of all possible states is $2^{M\times N\times P}$.\\
\begin{figure}[!hp]
\begin{center}
\includegraphics[width=\textwidth]{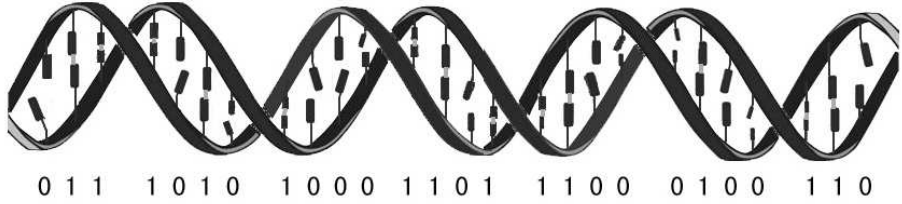}
\caption{A schematic graph for a double-stranded DNA molecule
associated with bool quantities to represent well-formed hydrogen
bonds (denoted as $1$) and decoupled bonds (denoted as $0$).}
\end{center}
\end{figure}

\indent The simulation begins at $t = 0$, $T =
0^{\,\circ}\textrm{C}$. At each step, $t$ increases by $1$, and
state of base pair at $(x_0, y_0, i)$ is inverted, i.e. $\psi(x_0,
y_0, i, t+1) = 1- (x_0, y_0, i, t)$. We assume that by changing
the state of the system for $M\times N\times P\times Z$ times, the
system will approximate the equilibrium state infinitely. The
change will be applied to each base pair for average $Z$ times.
$Z$ is determined by experience and should be reasonable. We
increase $T$ by  $\Delta T$ during the simulation. Therefore we
have the relation $T = \frac{t}{M\times N\times P\times Z}\Delta T$. \\
\indent Whether the state inversion is permitted depends on the
energy change ($\Delta E_t$) in each step. The possibility of the
state change at each step is
\begin{equation}
P(\psi(t)\rightarrow\psi(t+1))=\left\{
\begin{array}{c c c}
1 & for & \Delta E_t \leq0 \\
  &     &                  \\
e^{-\Delta E_t/k_BT} & for & \Delta E_t >0 \\
\end{array}
\right.
\end{equation}
\noindent If the current state change is permitted, we keep up
changing the system state at $t+1$. If the state change is
forbidden by the possibility, the system state remains unchanged
at $t$ and waits for another change at $t+1$. \\
\indent To achieve a relatively precise simulation, the change of
the total energy at time $t+1$ relative to that at the time $t$ is
analyzed by five components. The recursion relation of energy
change in each step is written as:
\begin{equation}
\Delta E_t = E(t+1) - E(t) = \sum_{l=1}^{5}\Delta E_l
\end{equation}
\noindent where E(t+1) and E(t) are the system energy for the
instant $t+1$ and $t$ respectively, and $\Delta E_l\,(l=1,2,3,4,5)$
is the variation of the $l$th component.\\
\indent The energy change depends on both the recursion relation
of the base pair at $(x_0, y_0, i)$ and the states of its nearest
neighbors. The global energy variation is determined by the local
states around the base pair $(x_0, y_0, i)$. Following analysis
presents the recursion relation of energy changes.\\

\noindent\textit{4.1. The hydrogen-binding energy ($\Delta E_1$)}\\
\noindent This component of the energy consists of the Morse
potential (equation (3)) and kinetic energy along the orientation
of the hydrogen bonds. The binding energy is independent of the
states of its neighboring base pairs.
\begin{equation}
\Delta E_1 = \left\{
\begin{array}{ccc}
  J & if & \psi(x_0, y_0, i, t)=0 \\
    &   &   \\
  -J & if & \psi(x_0, y_0, i, t)=1 \\
\end{array}
\right.
\end{equation}
\noindent where $J (J<0)$ is the binding energy for each base pair
that is in '1' state, while the binding energy for the '0' state
is zero
to be reference.\\

\noindent\textit{4.2. The stacking energy ($\Delta E_2$)}\\
\noindent To simplify the calculation of the stacking energy shown
in equation (2), we take into account the states of base pairs at
$(x_0, y_0, i-1, t)$ and $(x_0, y_0, i+1, t)$. Their states remain
unchanged during the interval from $t$ to $t+1$. We employ the
periodic boundary condition (PBC) listed below.
\begin{equation}
\begin{array}{c}
\psi(x+M,y,i,t)=\psi(x,y,i,t) \\
\psi(x,y+N,i,t)=\psi(x,y,i,t) \\
\psi(x,y,i+P,t)=\psi(x,y,i,t) \\
\end{array}
\end{equation}
\noindent Therefore $\psi(x_0, y_0, i-1, t)$ and $\psi(x_0, y_0,
i+1, t)$ are both well defined. The stacking energy reflects the
interaction between nearest neighboring base pairs in same DNA,
and it exists only when two nearest neighbors are in '1' state at
the same time. We use the symbol $\psi(x, y, {i_1, i_2,\ldots,
i_n}, t) = {b_1, b_2,\ldots, b_n}$ to denote states in the same
DNA for convenience, which means  $(x, y, i_1, t) = b_1$,  $(x, y,
i_2, t)=b_2$ ,\ldots,  $(x, y, i_n, t) = b_n$.
\begin{equation}
\Delta E_2 = \left\{
\begin{array}{ccc}
  0 & if & \psi(x_0, y_0, {i-1,i,i+1}, t)=\{000\} \\
  0 & if & \psi(x_0, y_0, {i-1,i,i+1}, t)=\{010\} \\
  w & if & \psi(x_0, y_0, {i-1,i,i+1}, t)=\{001\} \\
 -w & if & \psi(x_0, y_0, {i-1,i,i+1}, t)=\{011\} \\
  w & if & \psi(x_0, y_0, {i-1,i,i+1}, t)=\{100\} \\
 -w & if & \psi(x_0, y_0, {i-1,i,i+1}, t)=\{110\} \\
 2w & if & \psi(x_0, y_0, {i-1,i,i+1}, t)=\{101\} \\
-2w & if & \psi(x_0, y_0, {i-1,i,i+1}, t)=\{111\} \\
\end{array}
\right.
\end{equation}
\noindent where $w$ is the stacking energy stored in two nearest
neighboring
base pairs in '1' state.\\

\noindent\textit{4.3. Morse potential away from equilibrium point
($\Delta E_3$)}\\
\noindent We set $\Delta E_1 = 0$ for the uncoupled hydrogen bond
at base pairs. However, for a '0' state is next near to a '1'
state in the same DNA strand, the distance between two base pairs
is so close that the Morse potential should be taken into account.
We assigned energy $E$ to every two nearest neighboring base pairs
that are in different states in the same DNA.
\begin{equation}
\Delta E_3 = \left\{
\begin{array}{ccc}
 2E & if & \psi(x_0, y_0, {i-1,i,i+1}, t)=\{000\} \\
-2E & if & \psi(x_0, y_0, {i-1,i,i+1}, t)=\{010\} \\
  0 & if & \psi(x_0, y_0, {i-1,i,i+1}, t)=\{001\} \\
  0 & if & \psi(x_0, y_0, {i-1,i,i+1}, t)=\{011\} \\
  0 & if & \psi(x_0, y_0, {i-1,i,i+1}, t)=\{100\} \\
  0 & if & \psi(x_0, y_0, {i-1,i,i+1}, t)=\{110\} \\
-2E & if & \psi(x_0, y_0, {i-1,i,i+1}, t)=\{101\} \\
 2E & if & \psi(x_0, y_0, {i-1,i,i+1}, t)=\{111\} \\
\end{array}
\right.
\end{equation}\\

\noindent\textit{4.4. The effect of excluded volume ($\Delta E_4$)}\\
\noindent The effect of excluded volume in the nature of DNA phase
transition is discussed in Fisher's work [20]. The excluded volume
effect is connected to the system entropy variation. The effect is
prone to separate two complementary strands in a double helix. We
use $F$ to represent the energy change corresponding to this
effect. One should notice $\frac{\partial F}{\partial T}<0$. We
then have
\begin{equation}
\Delta E_4 = \left\{
\begin{array}{ccc}
 -F & if & \psi(x_0, y_0, {i-1,i,i+1}, t)=\{000\} \\
  F & if & \psi(x_0, y_0, {i-1,i,i+1}, t)=\{010\} \\
 -F & if & \psi(x_0, y_0, {i-1,i,i+1}, t)=\{001\} \\
  F & if & \psi(x_0, y_0, {i-1,i,i+1}, t)=\{011\} \\
 -F & if & \psi(x_0, y_0, {i-1,i,i+1}, t)=\{100\} \\
  F & if & \psi(x_0, y_0, {i-1,i,i+1}, t)=\{110\} \\
 -F & if & \psi(x_0, y_0, {i-1,i,i+1}, t)=\{101\} \\
  F & if & \psi(x_0, y_0, {i-1,i,i+1}, t)=\{111\} \\
\end{array}
\right.
\end{equation}
\indent The energy changes discussed above are summarized in the
figure 4 below, which does not take into account the DNA-DNA
interactions so far.
\begin{figure}[!htbp]
\begin{center}
\includegraphics[width=\textwidth]{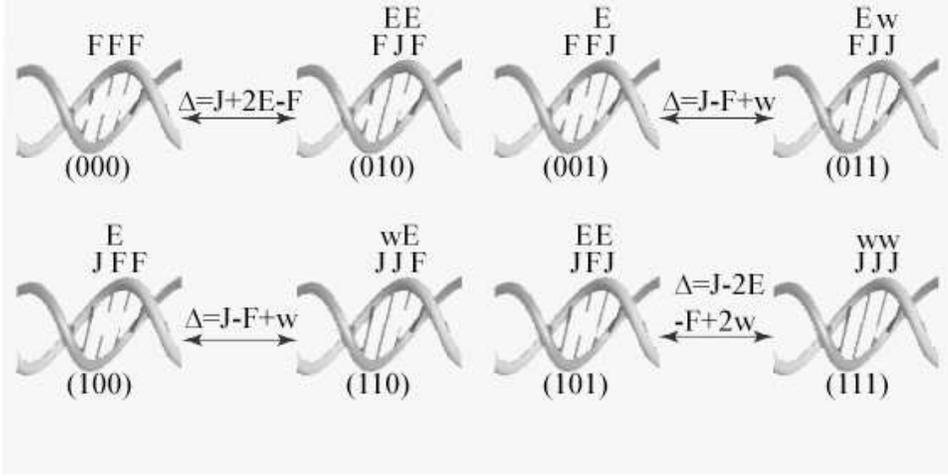}
\caption{Scheme for the energy changes. We calculate the energy
change under every case when the state of a base pair takes
inversion. The value overhanging the double-arrow represents the
energy variation when the state changes from left to right $\Delta
= \Delta E_1+\Delta E_2+\Delta E_3+\Delta E_4$ .}
\end{center}
\end{figure}

\noindent\textit{4.5. DNA-DNA interaction potential ($\Delta E_5$)}\\
\noindent We have introduced DNA-DNA interaction in previous
section. For each base pair, we denote the state of its $m$
nearest neighbors with $\lambda_i$,
($i=1,2,\ldots,m;\,\lambda_i=0,1$). $\Delta E_5$ can be written as
\begin{equation}
\Delta E_5 = \left\{
\begin{array}{ccc}
  G\sum_{i=1}^{m}\lambda_i & if & \psi(x_0, y_0, i, t)=0 \\
    &   &   \\
 -G\sum_{i=1}^{m}\lambda_i & if & \psi(x_0, y_0, i, t)=1 \\
\end{array}
\right.
\end{equation}
\noindent where $G$ is the interaction energy between each pair of
ions. Adding  $\Delta E5$ to $\Delta$, we will get the energy
variation
including the DNA-DNA interaction.\\

\noindent\textbf{5.Results and Discussions}\\
\noindent Following the equation (5) - (12), we could achieve a
coarse-gained simulation of the melting curves of ADNC as well as
that of DNA in bulk solutions. To perform the task, we choose
suitable scale parameters to carry out the simulation: $M = 100$,
$N = 100$, $P = 20$. The values of $M$ and $N$ chosen are much
smaller than ones of the actual situation, which is up to $10^{4}$
in the experiment. Since we take the periodic boundary condition,
the values of $M$ and $N$ used do not change our result. The
starting temperature is $0\,^{\circ}\textrm{C}$, and the final
temperature is $100\,^{\circ}\textrm{C}$, with increment of
$0.01\,^{\circ}\textrm{C}$ for each step. To guarantee the system
reaches equilibrium state, we take state changes under a specific
temperature. Each base pair has average $5$ times to be changed.
At each step, we count the number of DNAs that is still hybridized
and calculate the percentage for dsDNA in ADNC. The simulation
result shown in figure $5$ shows a steep melting transition
(hollow circles), consistent to the experimental observations. The
simulated result without considering DNA-DNA interaction show in
filled circles in figure $5$ also agrees with the DNA melting
curves in bulk solution. Comparison between the two cases suggests
that the DNA-DNA interaction greatly increases the melting point of
dsDNA chains.\\
\begin{figure}[!htb]
\begin{center}
\includegraphics[width=0.8\textwidth]{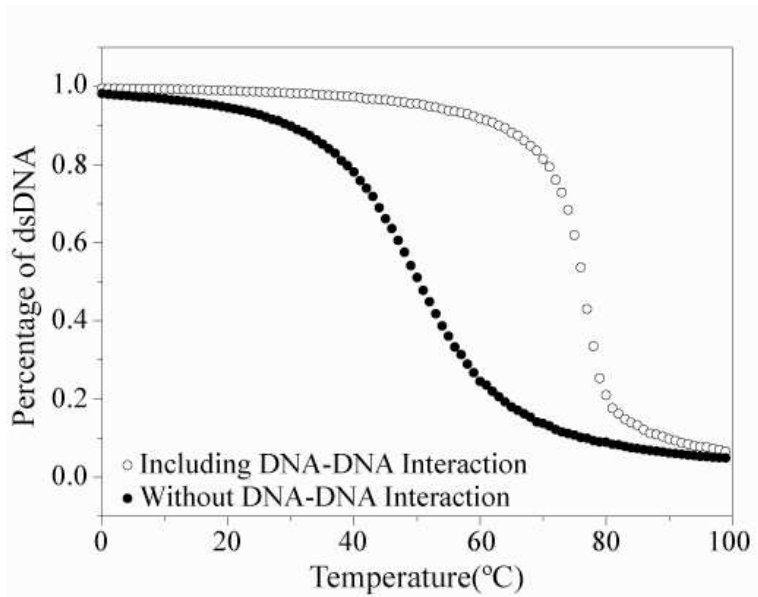}
\caption{Simulation results of melting curves for collective DNA
molecules. The filled circle represents the phase transition curve
without considering the DNA-DNA interaction, while the hollow
circle is the counterpart that takes into account the interaction.
Parameters used in the simulations are: $J = -1900k_B$, $w = -250
k_B$, $E = 850 k_B$, $F = -1650 k_B$, $\frac{\partial F}{\partial
T}= -10 k_B$, $G = -125 k_B$.}
\end{center}
\end{figure}

\indent In conclusion, we have established a simple coarse-gained
model to simulate the melting transition of DNA in ADNC. The
result provides a reasonable explanation for our experimental
observations. Although the simulation method discretizes the Morse
potential and stacking energy proposed in Peyrard-Bishop model,
the result still present a comparable approximation to
experimental data due to our fine treatment of energy changes
during melting transition. However, this work is only the
beginning of insightful theoretical investigation for the
rationality of ADNC. In future work, we will establish a more
precise model to employ an extensive investigation of phase
transition occurring in ADNC as well as its derived DNA
machines.\\

\noindent \textbf{Acknowledgements}\\
\noindent This work was partly supported by the grants from
Chinese Natural Science Foundation, Ministry of Science and
Technology of China and financial support from Peking
University.\\

\noindent\textbf{Reference}\\
$[1]$\quad\, Lifson S 1964 \textit{J. Chem. Phys.} \textbf{40}
3705\\
$[2]$\quad\, Poland D and Scheraga H A 1966 \textit{J. Chem.
Phys.} \textbf{45} 1456\\
$[3]$\quad\, Poland D and Scheraga A H 1966 \textit{J. Chem.
Phys.} \textbf{45} 1464.\\
$[4]$\quad\, Zhang Y L, Zheng W M, Liu J X, and Chen Y Z 1997
\textit{Phys. Rev.} E\\
\indent \quad \textbf{56} 7100\\
$[5]$\quad\, Theodorakopoulos N, Dauxois T and Peyard M 2000
\textit{Phys. Rev. Lett.}\\
\indent \quad \textbf{85} 6\\
$[6]$\quad\, Dauxois T and Peyard M 1995 \textit{Phys. Rev.} E
\textbf{51} 4027\\
$[7]$\quad\, Dauxois T and Peyard M 1993 \textit{Phys. Rev.} E
\textbf{47} R44\\
$[8]$\quad\, Mao Y D, Luo C X, Deng W, Jin G. Y, Yu X M, Zhang Z
H, Ouyang\\
\indent \quad Q, Chen R S and Yu D P 2004 \textit{Nucleic Acids
Res.} \textbf{32} e144\\
$[9]$\quad\, Mao Y D, Luo C X and Ouyang Q 2003 \textit{Nucleic
Acids Res.} \textbf{31} e108\\
$[10]$\quad Zimm B H and Bragg J K 1959 \textit{J. Chem. Phys.}
\textbf{28} 1246\\
$[11]$\quad Chrisey L A, Lee G U and O'Ferrall C E 1996
\textit{Nucleic Acids Res.} \textbf{24}\\
\indent \quad 3031\\
$[12]$\quad Bard A J and Fulkner L R 1980 \textit{Electrochemical
methods} Wiley, New\\
\indent \quad York\\
$[13]$\quad Harreis H M, Kornyshev A A, Likos C N, Lowen H, and
Sutmann G\\
\indent \quad 2002 \textit{Phys. Rev. Lett.} \textbf{89} 018303\\
$[14]$\quad Harreis H M, Likos C N, and Lowen H 2003
\textit{Biophys. J.} \textbf{84} 3607\\
$[15]$\quad Kornyshev A A and Leikin S 1997 \textit{J. Chem.
Phys.} \textbf{107} 3656\\
$[16]$\quad Kornyshev A A 2000 \textit{Phys. Rev.} E \textbf{62}
2576\\
$[17]$\quad Allahyarov E and Lowen H 2000 \textit{Phys. Rev.} E
\textbf{62} 5542\\
$[18]$\quad Kornyshev A A 2001 \textit{Phys. Rev. Lett.}
\textbf{86} 3666\\
$[19]$\quad Hill T L 1959 \textit{J. Chem. Phys.} \textbf{30}
383\\
$[20]$\quad Fisher M E 1966 \textit{J. Chem. Phys.} \textbf{45}
1469\\

\end{document}